\documentclass[aps,prl,reprint, superscriptaddress,floatfix,twocolumn,amsmath,amssymb]{revtex4-1}

\usepackage{graphicx}
\usepackage{dcolumn}
\usepackage{bm}
\usepackage{natbib}
\usepackage{newfloat}

\begin{document}


\title{Optomechanical Raman-Ratio Thermometry}


\author{T. P. Purdy}
\email[]{tpp@jila.colorado.edu}
\author{P.-L. Yu}
\author{N. S. Kampel}
\author{R. W. Peterson}
\affiliation{JILA, University of Colorado and National Institute of Standards and Technology,}
\affiliation{Department of Physics, University of Colorado, Boulder, CO 80309, USA}
\author{K. Cicak}
\author{R. W. Simmonds}
\affiliation{National Institute of Standards and Technology, Boulder, CO 80305, USA}
\author{C. A. Regal}
\email{regal@colorado.edu}
\affiliation{JILA, University of Colorado and National Institute of Standards and Technology,}
\affiliation{Department of Physics, University of Colorado, Boulder, CO 80309, USA}


\date{\today}

\begin{abstract}
	The temperature dependence of the asymmetry between Stokes and anti-Stokes Raman scattering can be exploited for self-calibrating, optically-based thermometry.  In the context of cavity optomechanics, we observe the cavity-enhanced scattering of light interacting with the standing-wave drumhead modes of a Si$_3$N$_4$ membrane mechanical resonator.  The ratio of the amplitude of Stokes to anti-Stokes scattered light is used to measure temperatures of optically-cooled mechanical modes down to the level of a few vibrational quanta.  We demonstrate that the Raman-ratio technique is able to measure the physical temperature of our device over a range extending from cryogenic temperatures to within an order of magnitude of room temperature.   
\end{abstract}


\maketitle

	Raman light scattering has proven to be a robust and powerful technique for in situ thermometry.  Many material-specific properties governing Raman transitions, such as the Stokes shift, spectral linewidth, and scattering rate vary with temperature.  However, for all Raman systems the ratio of spontaneously scattered Stokes versus anti-Stokes photons is a direct measure of the initial population of the motional state.  For example, at zero temperature the process of anti-Stokes scattering, which attempts to lower the motional state below the ground state, is entirely suppressed, whereas the Stokes scattering, which raises the motional state, is allowed. For thermally occupied states, an absolute, self-calibrating temperature measurement is possible by measuring this asymmetry in Raman scattering. Distributed optical fiber sensors~\cite{Dakin85} and solid state systems~\cite{Hart70,Kip90,Cui98} make use of spontaneous Raman scattering between optical phonon levels for temperature measurements, and combustion chemistry diagnostics use rotational-vibrational molecular levels in a similar fashion~\cite{Eckbreth96}.  Ultracold trapped ions~\cite{Diedrich89,Monroe95} and neutral atoms~\cite{Jessen92,Kaufman12} employ motional Raman sideband spectroscopy to reveal thermal occupations near the quantum ground state.  Recent experiments in the field of quantum cavity optomechanics~\cite{SafaviNaeini12,Brahms12,Weinstein14} use cavity enhancement to collect Raman-scattered light from localized acoustic resonances demonstrating the Stokes/anti-Stokes asymmetry.

\begin{figure}
	\centering
		\includegraphics[width=.9\linewidth]{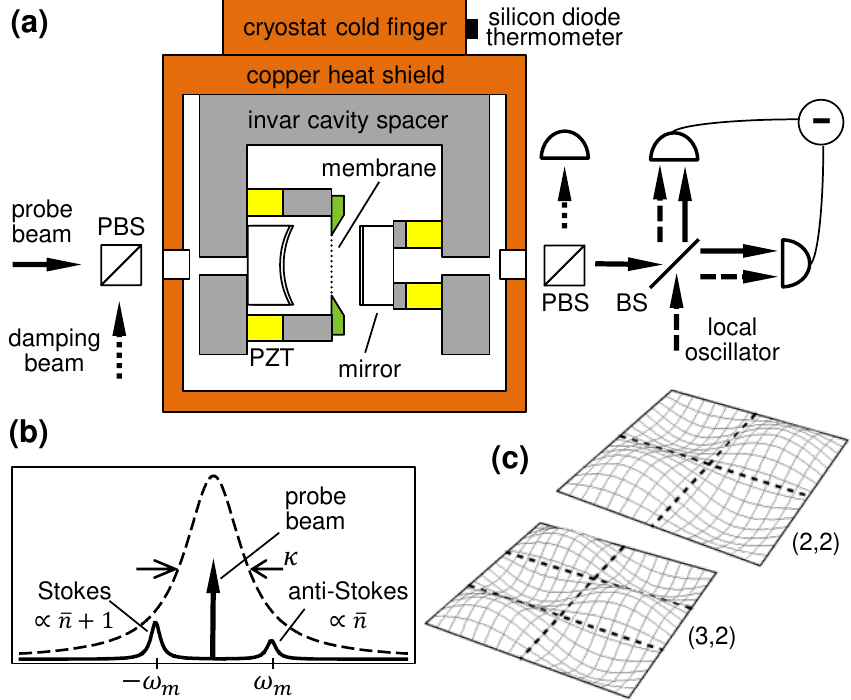}
		\caption{Optomechanical Raman-ratio thermometry. (a) Two orthogonally polarized laser beams are coupled into a cryogenic membrane-in-cavity optomechanical system.  Raman scattered light from the resonant probe beam is analyzed with balanced heterodyne detection.  The red detuned damping beam Raman sideband cools membrane motion. Beam splitter (BS), polarizing beam splitter (PBS), piezoelectric transducer (PZT). (b) The spectrum of transmitted probe beam light shows Raman scattering peaks shifted by $\pm \omega_m$, where $\omega_m$ is the mechanical resonance frequency.  The asymmetry in the spectral peaks, dependent on the effective thermal occupation of the mechanical mode, forms the basis for a self-calibrating thermometer.  Also indicated is the spectral response of the optical cavity (dashed curve).  (c) Spatial profile of the $(2,2)$ and $(3,2)$ drumhead mode of the membrane.}
	\label{fig:Figure1}
\end{figure}

	Here we measure the asymmetry of Raman scattering from a single, resonant laser tone driving a membrane-in-cavity optomechanical system (Fig.~\ref{fig:Figure1}).  The motional states are the MHz frequency vibrational levels of a membrane mechanical resonator, and an optical resonance is provided by an optical cavity surrounding the membrane.  The asymmetry becomes more pronounced in certain mechanical normal modes of the resonator when they are optically cooled near their ground state with a separate laser tone.  We use these measurements to verify that the damped displacement spectral density near the membrane resonance frequency is equal to that expected from a resonator occupied with $\bar{n}\sim 2$ vibrational quanta ($\sim$150~$\mu$K effective temperature).  Additionally, we measure the physical temperature of our device by extrapolating the Raman sideband asymmetry to zero optical damping.  These measurements agree at the ten percent level with those of a conventional thermometer monitoring the sample temperature over a wide temperature range.  Our result represents an absolute, self-calibrated thermometry of the physical temperature of the membrane.

	Raman-ratio thermometry is based on the idea that at a finite temperature the ground state manifold of the Raman levels is occupied according to a well understood statistical weighting.  The ratio of Stokes to anti-Stokes Raman transitions is equal to $R_{sa}=e^{\hbar \omega_m/k_b T}=(\bar{n}+1)/\bar{n}$, where $\omega_m$ is the mechanical resonance frequency and $T$ is the temperature \cite{Marquardt07,WilsonRae07}.  The spectrum of Raman scattered light transmitted through an optomechanical cavity (Fig.~\ref{fig:Figure1}(b)) is given by:
\begin{equation}
S(\omega)\propto\frac{\bar{n}}{\left(\frac{\Gamma_m}{2}\right)^2+(\omega_m-\omega)^2}+\frac{\bar{n}+1}{\left(\frac{\Gamma_m}{2}\right)^2+(\omega_m+\omega)^2}\nonumber
\label{eq:eq1}
\end{equation}
The first (second) term corresponds to anti-Stokes (Stokes) scattering peak shifted by $\omega_m$ ($-\omega_m$) from the input laser frequency, and $\omega$ is the frequency relative to the input laser frequency.  The peaks are broadened by the mechanical linewidth, $\Gamma_m$.  The laser is assumed to be resonant with the optical cavity, and the optical cavity linewidth is assumed to be much larger than $\Gamma_m$.  The mechanical occupation is $\bar{n}=\bar{n}_{th}+\bar{n}_{ba}$.  Here, $\bar{n}_{th}$ is the average thermal occupation of the mechanical mode, which may be optically cooled well below the physical temperature of the system \cite{Marquardt07,WilsonRae07}.  $\bar{n}_{ba}$ is the mechanical backaction induced by radiation pressure from shot noise intensity fluctuations of the incident laser~\cite{Purdy13}, which can be made small compared to $\bar{n}_{th}$.  Taking the ratio of the amplitude of the Stokes and anti-Stokes peaks directly yields the mechanical occupation, $\bar{n}=1/(R_{sa}-1)$, without the need for additional input parameters or calibrations.

\begin{figure}
	\centering
		\includegraphics[width=\linewidth]{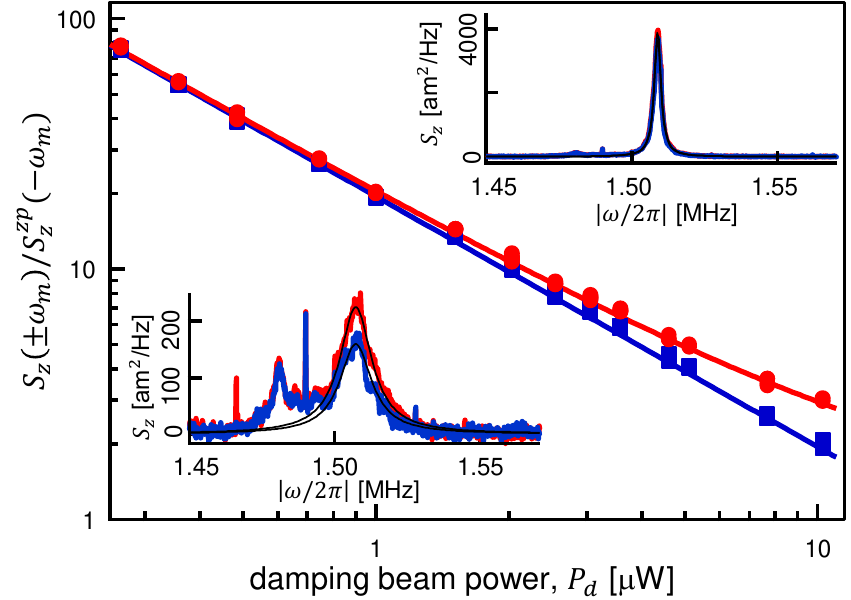}
		\caption{Peak amplitude of Stokes $S_{z}(-\omega_m)$ (red circles) and anti-Stokes $S_{z}(+\omega_m)$ (blue squares) Raman peaks of the $(2,2)$ mode. Spectral densities are expressed in units of mode displacement. The scale $S_{z}^{zp}(-\omega_m)=S_{z}(-\omega_m)|_{\bar{n}=0}$ corresponds to one quanta of vibrational motion. $P_p=5$~$\mu$W and $\kappa/2\pi=1.8$~MHz.  Raman spectra for $P_d=1.5$~ $\mu$W (7.7~$\mu$W) are shown in the upper (lower) inset, along with Lorentzian fits (including only the frequency range above $\sim$1.50~MHz) in black. Noise from thermally occupied mechanical modes of the substrate is visible below 1.50~MHz.  The white, detection-noise background is subtracted from the traces.}
	\label{fig:Figure2}
\end{figure}

\begin{figure}
	\centering
		\includegraphics[width=\linewidth]{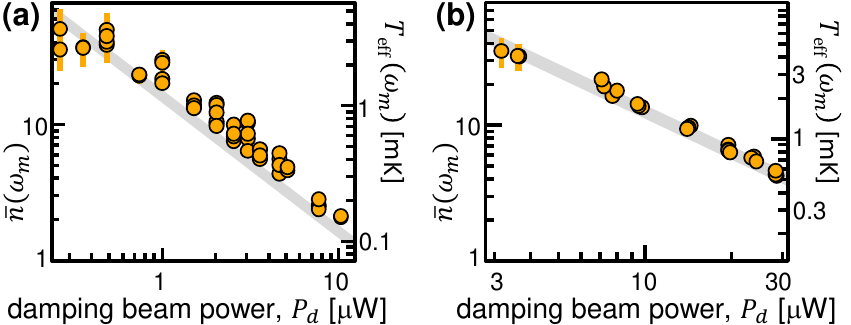}
		\caption{Effective mode occupation at mechanical resonance.  Numerically integrating the Raman peaks over a 4~kHz span around the mechanical resonance and taking their ratio yields a measure of the mode effective temperature (orange circles) for (a) the (2,2) mode of the non-phononic-crystal device and  (b) the (3,2) mode of the phononic-crystal device.  Due to decreased optomechanical coupling, the (3,2) mode does not reach as low a mechanical occupation as the (2,2) mode.  Gray bands are the prediction of Raman sideband cooling whose width stems from uncertainty in $T_0$ as measured by the diode thermometer.  Vertical error bars represent estimated statistical standard deviation.}
	\label{fig:Figure2p5}
\end{figure}

\begin{figure*}
	\centering
		\includegraphics[width=\linewidth]{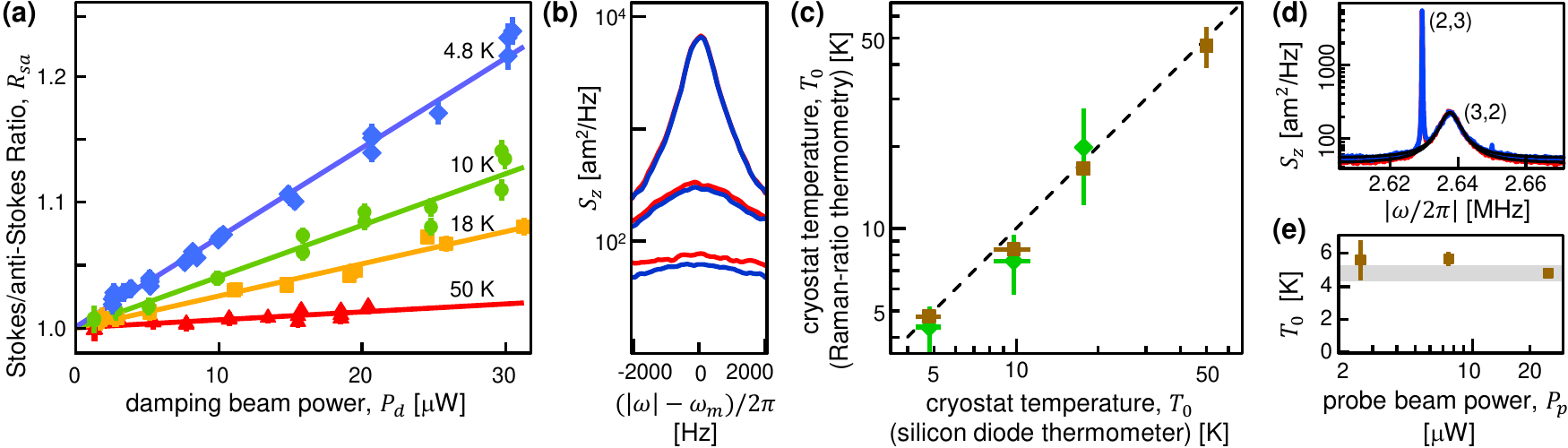}
		\caption{(a) Stokes/anti-Stokes peak ratio for the $(3,2)$ mode at four cryostat temperatures.  Solid curves are linear fits to the data.  $P_p=26$~$\mu$W and $\kappa/2\pi=2.7$~MHz.  (b) Stokes (red) and anti-Stokes (blue) spectra near the (3,2) mode for $P_d$ of 3.2~$\mu$W (top), 14~$\mu$W (middle), and 28~$\mu$W (bottom). Resolution bandwidth is 100~Hz.  (c) Comparison of Raman-ratio thermometry with silicon diode thermometer for the $(3,2)$ mode (brown squares) and $(5,2)$ mode (green diamonds).  Dashed line indicates agreement between the two methods.  (d) Raman spectra of (2,3) and (3,2) mode at $P_d=25$~$\mu$W.  Black curves are fit Lorentzians excluding the region around the (2,3) mode. (e) Extracted temperature, $T_0$, for various probe powers.  Vertical error bars in (a), (c), and (e) represent estimated statistical standard deviation.  Horizontal error bars in (c) and width of gray band in (e) represent a combination of the diode thermometer accuracy ($\pm 0.5$~K) and inaccuracy introduced by drift and oscillations in the temperature over the span of the measurement.}
	\label{fig:Figure3}
\end{figure*}

	Our membrane-in-cavity optomechanical system~\cite{Thompson08} consists of a high-tensile-stress, silicon nitride membrane in the standing wave of a Fabry-Perot optical resonator~\cite{Purdy12} (Fig.~\ref{fig:Figure1}(a)). The system is operated in a helium flow cryostat, with a temperature, $T_0$, that is set in the range of 4.8 to 50~K.  The mechanical modes that couple to the optical resonance are those of a square drum described by mode indicies $(m,n)$, which count the number of antinodes along each axis of the square.  We have employed two devices for our measurements.  The data shown in Figs.~\ref{fig:Figure2} and \ref{fig:Figure2p5}(a) focus on the $(2,2)$ mode of a 500~$\mu$m square by 40~nm thick membrane, with resonance frequency $\omega^{\scriptscriptstyle (2,2)}_m/2\pi=1.509$~MHz and intrinsic linewidth $\Gamma^{\scriptscriptstyle (2,2)}_0/2\pi=0.46$~Hz.  The data shown in Figs.~\ref{fig:Figure2p5}(b) and \ref{fig:Figure3} are primarily from the $(3,2)$ mode of a 375~$\mu$m square by 100~nm thick membrane, $\omega_m^{\scriptscriptstyle (3,2)}/2\pi=2.637$~MHz, $\Gamma^{\scriptscriptstyle (3,2)}_0/2\pi=0.84$~Hz, which is supported by a silicon substrate patterned into a square-lattice phononic crystal (device A of Ref.~\cite*{[{}] [{ We note that measured mechanical resonances of all of the membrane modes are shifted lower in frequency compared to those reported in this reference.  This shift moves the (3,2) resonance closer to the center of mechanical band gap.  The change may be the result of a stress reduction due to differences in sample mounting technique, mass loading from surface contamination, or device aging.}] Yu14}).  The acoustic band structure is engineered to provide a gap in the substrate mechanical mode density around the $(3,2)$ mode, which diminishes noise from thermally occupied modes of the substrate~\cite{Yu14,Tsaturyan14}.

	The optical cavity consists of two mirrors separated by 3.5 mm, each with $10^{-4}$ fractional transmission.  The optical linewidth, $\kappa$, is dependent on the location of the membrane in the cavity~\cite{Wilson09}, and is on the order of a few MHz.  Optomechanical coupling is achieved as the optical resonance frequency is modulated by the displacement of the vibrating membrane along the optical standing wave.  This interaction is characterized by a single photon coupling rate of $g_0^{\scriptscriptstyle (2,2)}/2\pi=33$~Hz for the data in Fig.~\ref{fig:Figure2} or $g_0^{\scriptscriptstyle (3,2)}/2\pi=18$~Hz for the data in Fig.~\ref{fig:Figure3}. The cavity is driven with two orthogonally polarized laser beams derived from the same 1064 nm laser source (Fig.~\ref{fig:Figure1}(a)).  The probe beam is actively stabilized to be resonant with the optical cavity.  The transmitted Raman scattered light from this beam is analyzed with an optical heterodyne detection system.  The orthogonal polarization mode is driven by the damping beam, which is tuned to a frequency lower than the optical resonance.

	By varying the power of the damping beam, we are able to damp motion and reduce the effective mode temperature by up to a factor of $3\times 10^4$ via Raman sideband cooling.  For the red-detuned damping beam, the anti-Stokes scattering rate is resonantly enhanced by the cavity~\cite{Marquardt07,WilsonRae07}.  Each anti-Stokes scattered photon that exits the cavity carries one vibrational quantum of energy out of the system.  The mode reaches an equilibrium when the optical cooling rate is matched by the rate at which thermal excitations enter the system.  The effective temperature of a mode is approximately $T_{\mathrm{eff}}=T_0 \frac{\Gamma_0}{\Gamma_m}$, in the experimentally relevant case where $\Gamma_0\ll \Gamma_m \ll \kappa$, here $\Gamma_m$ is the optically induced mechanical damping rate, and $\Gamma_0$ is the intrinsic mechanical damping rate.  Figure \ref{fig:Figure2} shows the amplitude of the Raman peaks generated by the resonant probe beam, which is held at constant transmitted power, $P_p$, as the damping beam power, $P_d$, is increased.  As expected, the height of the anti-Stokes peak decreases with damping beam power as the mode is cooled.  For weak damping the two Raman peaks are equal in amplitude, whereas at strongest damping the Stokes peak is 50\% larger than the anti-Stokes peak as the mode approaches its quantum ground state.

	From the Stokes/anti-Stokes ratio we compute the effective temperature of the mode, as displayed in Fig.~\ref{fig:Figure2p5}.  Raman-ratio thermometry indicates that for frequencies near $\omega_m$, the (2,2) mode reaches $\bar{n}(\omega_m)=2.1\pm0.2$.  However, for this mode, the occupations measured over a wide range of damping beam power are systematically $\sim$15\% larger than predicted by Raman sideband cooling.  This discrepency, which is absent in the phononic-crystal-isolated (3,2) mode (Fig.~\ref{fig:Figure2p5}(b)), is likely due to the influence of substrate motion and is discussed below.

	We next consider Raman-ratio thermometry as a means to determine the physical temperature of the device.  Here we employ a membrane resonator embedded in a phononic crystal substrate, which reduces the effects of substrate motion to a level below the statistical noise.  The Raman asymmetry of undamped modes is small ($\sim 10^{-4}$) and precludes directly ascertaining the physical device temperature.  Instead, we extrapolate the temperature of the undamped membrane modes from measurements of the asymmetry as a function of optical damping (Fig.~\ref{fig:Figure3}(a)).  The undamped occupation of the mechanical mode is given by $(dR_{sa}/d P_d)^{-1} \times d\Gamma_m/dP_d \times 1/\Gamma_0$.
	
	Thermometry is performed on the $(3,2)$ mode with the cryostat held at several different temperatures between 4.8~K and 50~K.   In Fig.~\ref{fig:Figure3}(c) the extrapolated temperatures are compared to the temperature as measured by a silicon diode thermometer attached to the cryostat.  The results agree within the statistical uncertainty with an average deviation of less than 10\%.  As an additional confirmation, we perform the same thermometry procedure on the $(5,2)$ mode, which is also located in a phononic band gap of the substrate and find agreement with the previous measurements.  However the statistical error on the $(5,2)$ mode measurement is much larger because the optomechanical coupling is smaller than to the $(3,2)$ mode.

	  Thermometry relying on sideband asymmetry does not require accurate knowledge of often-difficult-to-measure system parameters such as the optomechanical coupling rate, effective modal mass, circulating optical power, or optical detection efficiency.  Additionally, the single-resonant-probe method~\cite{Brahms12} employed in this work reduces the sensitivity to many systematic errors as compared to techniques that employ multiple off-resonant probe tones~\cite{SafaviNaeini12,Weinstein14}.  Such effects include drift in the relative amplitude of the probes, optomechanical modification of the mechanical susceptibility, coherent optomechanically induced interference between the probes~\cite{Weis10}, and increased sensitivity to classical laser noise~\cite{Jayich12}.  However, many systematic effects in our system can lead to error in temperature determination, and we now discuss several of these potential error sources.
	  
	A necessary external input parameter to extract the physical temperature is the intrinsic mechanical linewidth, which is measured via mechanical ringdown with an uncertainty of a few tenths of a percent.  However, long term drift in the mechanical linewidth can be at the percent level.  The intrinsic linewidth is recorded at each cryostat temperature, and found to increase from 0.84~Hz at 4.8~K to 1.07~Hz at 50~K.  The optically-damped mechanical linewidths are obtained by Lorentzian fits to Raman spectra.
	
	We measure the frequency dependence in the heterodyne detection system by weakly modulating the input probe beam at a frequency near the mechanical resonance. The ratio of detected modulation sidebands gives the heterodyne response. This calibration agrees at the percent level with the observed ratio of the Raman sidebands for weak optical damping.  So to correct for this frequency dispersion, we scale our data such that $R_{sa}=1$ in the absence of mechanical damping (i.~e.~the intercepts of the fits in Fig.~\ref{fig:Figure3}(a) are set to one).
	 
	Noise from thermally-occupied mechanical modes of the substrate is observed in the spectra of the non-phononic-crystal device (Fig.~\ref{fig:Figure2} inset, e.g. 1.47 to 1.50 MHz)~\cite{Yu14,Purdy12}.  Because the probe beam is tuned to the optical resonance, the noise is added to both laser sidebands.  We find the noise spectra to be approximately independent of damping beam power. To minimize the effect of substrate noise on thermometry, we have generated the data of Fig.~\ref{fig:Figure2p5} by integrating the Raman peaks in a narrow band ($\pm$2~kHz) around the mechanical resonance.  Integrating the excess noise over the entire span of Fig.~\ref{fig:Figure2} inset results in slightly less than one quantum of additional motion. We assess the systematic error from substrate motion near mechanical resonance by varying the center of the region of numerical integration and find that the estimated temperature can vary by a few tens of percent over a few kHz for this device.  For the phononic-crystal-substrate device (Fig.~\ref{fig:Figure3}(d)), this effect is mitigated below the statistical noise level by the cleaner spectrum provided by the acoustic band gap.
	 
	Heating induced by absorbed laser light can cause thermal gradients between the membrane and silicon diode thermometer and increase $\bar{n}_{th}$.  We see no evidence for absorptive heating, which would cause deviations from linearity in the data of Fig.~\ref{fig:Figure3}(a) and from the power law fit of the anti-Stokes data of Fig.~\ref{fig:Figure2}(a).  Another, more fundamental, systematic effect is motion driven by optical forces from intensity fluctuations on the probe beam, which increases $\bar{n}_{ba}$.  Because the noise on the probe beam is dominated by shot noise, the optical force fluctuations can be calculated as in Ref.~\cite{Purdy13}.  For $P_p=26$~$\mu$W probing the $(3,2)$ mode, the correction to $T_0$ is 0.2~K due to this effect.  This correction is applied to the extracted physical temperature data in Fig.~\ref{fig:Figure3}(c) and the theoretical expectation bands of Fig.~\ref{fig:Figure2p5}.
	  
	In addition to the quantum noise of the probe beam, optomechanically mediated correlations in the classical laser noise can also lead to systematic error.  Such noise-squashing or noise-cancellation effects can change the relative heights of the Raman peaks leading to an underestimate of the effective mode occupation~\cite{Jayich12,Pontin14,Weinstein14}.  To assess this systematic, we have independently measured the classical phase and amplitude noise as well as the detuning of the input probe beam.  The classical amplitude noise is less than 1\% of shot noise, and the phase noise is less than 5\% of shot noise for $P_p=26$~$\mu$W.  Phase noise measurements are performed using heterodyne detection of input light that is passed through a fiber optic delay line.  The detuning of the probe beam is estimated from its residual optical damping rate to be about -0.0015~$\kappa$ with a long term drift of around 0.001~$\kappa$.  These measurements constrain the potential error~\cite{Jayich12} on temperature due to classical laser noise correlations to be at the percent level or less under all operating conditions.  Additionally, we have reduced the probe beam power by an order of magnitude and observed only a weak trend in extracted temperature (Fig.~\ref{fig:Figure3}(e)), which confirms that the effects of both classical laser noise and absorptive heating are small.

	The $(3,2)$ membrane mode is nearly degenerate with the $(2,3)$ mode, with a frequency difference between the two modes of 7.2~kHz (Fig.~\ref{fig:Figure3}(d)).  These modes can couple through the optomechanical interaction if their optomechanical responses are comparable in magnitude~\cite{Shkarin14}.  However, the optomechanical coupling rate $g_0^{\scriptscriptstyle (3,2)}$ is four times larger than $g_0^{\scriptscriptstyle (2,3)}$, due to the poor spatial overlap of the optical mode spot with the $(2,3)$ mode.  For all values of employed optical damping the optomechanical response of the $(3,2)$ mode near $\omega_m^{\scriptscriptstyle (3,2)}$, $|g_0^{\scriptscriptstyle (3,2)}\chi_m^{\scriptscriptstyle (3,2)}(\omega_m^{\scriptscriptstyle (3,2)})|^2$, is well over an order of magnitude greater than that of the $(2,3)$ mode at the same frequency, $|g_0^{\scriptscriptstyle (2,3)}\chi_m^{\scriptscriptstyle (2,3)}(\omega_m^{\scriptscriptstyle (3,2)})|^2$, where $\chi_m^{\scriptscriptstyle (m,n)}(\omega)=(\Gamma_m^{\scriptscriptstyle (m,n)}/2-i(\omega-\omega_m^{\scriptscriptstyle (m,n)}))^{-1}$ is the mechanical susceptibility of the optically damped $(m,n)$ mode. For thermometry, Raman peaks are always compared within $\pm$2~kHz of the peak of $(3,2)$ mode resonance.  Thus we believe the presence of the $(2,3)$ mode is a negligible perturbation on the thermometry data shown in Fig.~\ref{fig:Figure3}.  Increasing the optical damping well beyond the level presented here could cause mode hybridization and significantly complicate the interpretation of the Raman spectra~\cite{Shkarin14}.


	In conclusion, we have shown that Raman asymmetry is a viable technique to diagnose both the effective mode temperature of an optically damped membrane resonator and the physical temperature of the same device.  These measurements demonstrate the self-calibrating nature of the method and elucidate many of the systematic uncertainties.  Our results show that the quantum effects governing the asymmetry~\cite{Khalili12,Weinstein14} are visible in a membrane-in-cavity optomechanical system operated within an order of magnitude of room temperature.

\begin{acknowledgments}
 In the course of this work we became aware of parallel studies~\cite{Lee14}.

 This work was supported by the DARPA QuASAR program, the ONR YIP, the AFOSR, and the National Science Foundation under grant number 1125844. C.R. thanks the Clare Boothe Luce Foundation for support. P.-L.Y. thanks the Taiwan Ministry of Education for support.

\end{acknowledgments}


\begin{thebibliography}{26}%
\makeatletter
\providecommand \@ifxundefined [1]{%
 \@ifx{#1\undefined}
}%
\providecommand \@ifnum [1]{%
 \ifnum #1\expandafter \@firstoftwo
 \else \expandafter \@secondoftwo
 \fi
}%
\providecommand \@ifx [1]{%
 \ifx #1\expandafter \@firstoftwo
 \else \expandafter \@secondoftwo
 \fi
}%
\providecommand \natexlab [1]{#1}%
\providecommand \enquote  [1]{``#1''}%
\providecommand \bibnamefont  [1]{#1}%
\providecommand \bibfnamefont [1]{#1}%
\providecommand \citenamefont [1]{#1}%
\providecommand \href@noop [0]{\@secondoftwo}%
\providecommand \href [0]{\begingroup \@sanitize@url \@href}%
\providecommand \@href[1]{\@@startlink{#1}\@@href}%
\providecommand \@@href[1]{\endgroup#1\@@endlink}%
\providecommand \@sanitize@url [0]{\catcode `\\12\catcode `\$12\catcode
  `\&12\catcode `\#12\catcode `\^12\catcode `\_12\catcode `\%12\relax}%
\providecommand \@@startlink[1]{}%
\providecommand \@@endlink[0]{}%
\providecommand \url  [0]{\begingroup\@sanitize@url \@url }%
\providecommand \@url [1]{\endgroup\@href {#1}{\urlprefix }}%
\providecommand \urlprefix  [0]{URL }%
\providecommand \Eprint [0]{\href }%
\providecommand \doibase [0]{http://dx.doi.org/}%
\providecommand \selectlanguage [0]{\@gobble}%
\providecommand \bibinfo  [0]{\@secondoftwo}%
\providecommand \bibfield  [0]{\@secondoftwo}%
\providecommand \translation [1]{[#1]}%
\providecommand \BibitemOpen [0]{}%
\providecommand \bibitemStop [0]{}%
\providecommand \bibitemNoStop [0]{.\EOS\space}%
\providecommand \EOS [0]{\spacefactor3000\relax}%
\providecommand \BibitemShut  [1]{\csname bibitem#1\endcsname}%
\let\auto@bib@innerbib\@empty
\bibitem [{\citenamefont {Dakin}\ \emph {et~al.}(1985)\citenamefont {Dakin},
  \citenamefont {Pratt}, \citenamefont {Ross},\ and\ \citenamefont
  {Bibby}}]{Dakin85}%
  \BibitemOpen
  \bibfield  {author} {\bibinfo {author} {\bibfnamefont {J.~P.}\ \bibnamefont
  {Dakin}}, \bibinfo {author} {\bibfnamefont {D.~J.}\ \bibnamefont {Pratt}},
  \bibinfo {author} {\bibfnamefont {J.~N.}\ \bibnamefont {Ross}}, \ and\
  \bibinfo {author} {\bibfnamefont {G.~W.}\ \bibnamefont {Bibby}},\ }in\
  \href@noop {} {\emph {\bibinfo {booktitle} {Proc. Conf. on Optical-Fiber
  Sensors 3}}}\ (\bibinfo {address} {San Diego},\ \bibinfo {year}
  {1985})\BibitemShut {NoStop}%
\bibitem [{\citenamefont {Hart}\ \emph {et~al.}(1970)\citenamefont {Hart},
  \citenamefont {Aggarwal},\ and\ \citenamefont {Lax}}]{Hart70}%
  \BibitemOpen
  \bibfield  {author} {\bibinfo {author} {\bibfnamefont {T.~R.}\ \bibnamefont
  {Hart}}, \bibinfo {author} {\bibfnamefont {R.~L.}\ \bibnamefont {Aggarwal}},
  \ and\ \bibinfo {author} {\bibfnamefont {B.}~\bibnamefont {Lax}},\ }\href
  {\doibase 10.1103/PhysRevB.1.638} {\bibfield  {journal} {\bibinfo  {journal}
  {Phys. Rev. B}\ }\textbf {\bibinfo {volume} {1}},\ \bibinfo {pages} {638}
  (\bibinfo {year} {1970})}\BibitemShut {NoStop}%
\bibitem [{\citenamefont {Kip}\ and\ \citenamefont {Meier}(1990)}]{Kip90}%
  \BibitemOpen
  \bibfield  {author} {\bibinfo {author} {\bibfnamefont {B.~J.}\ \bibnamefont
  {Kip}}\ and\ \bibinfo {author} {\bibfnamefont {R.~J.}\ \bibnamefont
  {Meier}},\ }\href
  {http://www.opticsinfobase.org/as/abstract.cfm?uri=as-44-4-707} {\bibfield
  {journal} {\bibinfo  {journal} {Appl. Spectrosc.}\ }\textbf {\bibinfo
  {volume} {44}},\ \bibinfo {pages} {707} (\bibinfo {year} {1990})}\BibitemShut
  {NoStop}%
\bibitem [{\citenamefont {Cui}\ \emph {et~al.}(1998)\citenamefont {Cui},
  \citenamefont {Amtmann}, \citenamefont {Ristein},\ and\ \citenamefont
  {Ley}}]{Cui98}%
  \BibitemOpen
  \bibfield  {author} {\bibinfo {author} {\bibfnamefont {J.~B.}\ \bibnamefont
  {Cui}}, \bibinfo {author} {\bibfnamefont {K.}~\bibnamefont {Amtmann}},
  \bibinfo {author} {\bibfnamefont {J.}~\bibnamefont {Ristein}}, \ and\
  \bibinfo {author} {\bibfnamefont {L.}~\bibnamefont {Ley}},\ }\href {\doibase
  10.1063/1.367972} {\bibfield  {journal} {\bibinfo  {journal} {Journal of
  Applied Physics}\ }\textbf {\bibinfo {volume} {83}},\ \bibinfo {pages} {7929}
  (\bibinfo {year} {1998})}\BibitemShut {NoStop}%
\bibitem [{\citenamefont {Eckbreth}(1996)}]{Eckbreth96}%
  \BibitemOpen
  \bibfield  {author} {\bibinfo {author} {\bibfnamefont {A.~C.}\ \bibnamefont
  {Eckbreth}},\ }\href@noop {} {\emph {\bibinfo {title} {Laser Diagnostics for
  Combustion Temperature and Species}}},\ \bibinfo {edition} {2nd}\ ed.\
  (\bibinfo  {publisher} {Gordon and Breach},\ \bibinfo {address} {Amsterdam,
  Netherlands},\ \bibinfo {year} {1996})\ Chap.~\bibinfo {chapter}
  {5}\BibitemShut {NoStop}%
\bibitem [{\citenamefont {Diedrich}\ \emph {et~al.}(1989)\citenamefont
  {Diedrich}, \citenamefont {Bergquist}, \citenamefont {Itano},\ and\
  \citenamefont {Wineland}}]{Diedrich89}%
  \BibitemOpen
  \bibfield  {author} {\bibinfo {author} {\bibfnamefont {F.}~\bibnamefont
  {Diedrich}}, \bibinfo {author} {\bibfnamefont {J.~C.}\ \bibnamefont
  {Bergquist}}, \bibinfo {author} {\bibfnamefont {W.~M.}\ \bibnamefont
  {Itano}}, \ and\ \bibinfo {author} {\bibfnamefont {D.~J.}\ \bibnamefont
  {Wineland}},\ }\href {\doibase 10.1103/PhysRevLett.62.403} {\bibfield
  {journal} {\bibinfo  {journal} {Phys. Rev. Lett.}\ }\textbf {\bibinfo
  {volume} {62}},\ \bibinfo {pages} {403} (\bibinfo {year} {1989})}\BibitemShut
  {NoStop}%
\bibitem [{\citenamefont {Monroe}\ \emph {et~al.}(1995)\citenamefont {Monroe},
  \citenamefont {Meekhof}, \citenamefont {King}, \citenamefont {Jefferts},
  \citenamefont {Itano}, \citenamefont {Wineland},\ and\ \citenamefont
  {Gould}}]{Monroe95}%
  \BibitemOpen
  \bibfield  {author} {\bibinfo {author} {\bibfnamefont {C.}~\bibnamefont
  {Monroe}}, \bibinfo {author} {\bibfnamefont {D.~M.}\ \bibnamefont {Meekhof}},
  \bibinfo {author} {\bibfnamefont {B.~E.}\ \bibnamefont {King}}, \bibinfo
  {author} {\bibfnamefont {S.~R.}\ \bibnamefont {Jefferts}}, \bibinfo {author}
  {\bibfnamefont {W.~M.}\ \bibnamefont {Itano}}, \bibinfo {author}
  {\bibfnamefont {D.~J.}\ \bibnamefont {Wineland}}, \ and\ \bibinfo {author}
  {\bibfnamefont {P.}~\bibnamefont {Gould}},\ }\href {\doibase
  10.1103/PhysRevLett.75.4011} {\bibfield  {journal} {\bibinfo  {journal}
  {Phys. Rev. Lett.}\ }\textbf {\bibinfo {volume} {75}},\ \bibinfo {pages}
  {4011} (\bibinfo {year} {1995})}\BibitemShut {NoStop}%
\bibitem [{\citenamefont {Jessen}\ \emph {et~al.}(1992)\citenamefont {Jessen},
  \citenamefont {Gerz}, \citenamefont {Lett}, \citenamefont {Phillips},
  \citenamefont {Rolston}, \citenamefont {Spreeuw},\ and\ \citenamefont
  {Westbrook}}]{Jessen92}%
  \BibitemOpen
  \bibfield  {author} {\bibinfo {author} {\bibfnamefont {P.~S.}\ \bibnamefont
  {Jessen}}, \bibinfo {author} {\bibfnamefont {C.}~\bibnamefont {Gerz}},
  \bibinfo {author} {\bibfnamefont {P.~D.}\ \bibnamefont {Lett}}, \bibinfo
  {author} {\bibfnamefont {W.~D.}\ \bibnamefont {Phillips}}, \bibinfo {author}
  {\bibfnamefont {S.~L.}\ \bibnamefont {Rolston}}, \bibinfo {author}
  {\bibfnamefont {R.~J.~C.}\ \bibnamefont {Spreeuw}}, \ and\ \bibinfo {author}
  {\bibfnamefont {C.~I.}\ \bibnamefont {Westbrook}},\ }\href {\doibase
  10.1103/PhysRevLett.69.49} {\bibfield  {journal} {\bibinfo  {journal} {Phys.
  Rev. Lett.}\ }\textbf {\bibinfo {volume} {69}},\ \bibinfo {pages} {49}
  (\bibinfo {year} {1992})}\BibitemShut {NoStop}%
\bibitem [{\citenamefont {Kaufman}\ \emph {et~al.}(2012)\citenamefont
  {Kaufman}, \citenamefont {Lester},\ and\ \citenamefont {Regal}}]{Kaufman12}%
  \BibitemOpen
  \bibfield  {author} {\bibinfo {author} {\bibfnamefont {A.~M.}\ \bibnamefont
  {Kaufman}}, \bibinfo {author} {\bibfnamefont {B.~J.}\ \bibnamefont {Lester}},
  \ and\ \bibinfo {author} {\bibfnamefont {C.~A.}\ \bibnamefont {Regal}},\
  }\href {\doibase 10.1103/PhysRevX.2.041014} {\bibfield  {journal} {\bibinfo
  {journal} {Phys. Rev. X}\ }\textbf {\bibinfo {volume} {2}},\ \bibinfo {pages}
  {041014} (\bibinfo {year} {2012})}\BibitemShut {NoStop}%
\bibitem [{\citenamefont {Safavi-Naeini}\ \emph {et~al.}(2012)\citenamefont
  {Safavi-Naeini}, \citenamefont {Chan}, \citenamefont {Hill}, \citenamefont
  {Alegre}, \citenamefont {Krause},\ and\ \citenamefont
  {Painter}}]{SafaviNaeini12}%
  \BibitemOpen
  \bibfield  {author} {\bibinfo {author} {\bibfnamefont {A.~H.}\ \bibnamefont
  {Safavi-Naeini}}, \bibinfo {author} {\bibfnamefont {J.}~\bibnamefont {Chan}},
  \bibinfo {author} {\bibfnamefont {J.~T.}\ \bibnamefont {Hill}}, \bibinfo
  {author} {\bibfnamefont {T.~P.~M.}\ \bibnamefont {Alegre}}, \bibinfo {author}
  {\bibfnamefont {A.}~\bibnamefont {Krause}}, \ and\ \bibinfo {author}
  {\bibfnamefont {O.}~\bibnamefont {Painter}},\ }\href {\doibase
  10.1103/PhysRevLett.108.033602} {\bibfield  {journal} {\bibinfo  {journal}
  {Phys. Rev. Lett.}\ }\textbf {\bibinfo {volume} {108}},\ \bibinfo {pages}
  {033602} (\bibinfo {year} {2012})}\BibitemShut {NoStop}%
\bibitem [{\citenamefont {Brahms}\ \emph {et~al.}(2012)\citenamefont {Brahms},
  \citenamefont {Botter}, \citenamefont {Schreppler}, \citenamefont {Brooks},\
  and\ \citenamefont {Stamper-Kurn}}]{Brahms12}%
  \BibitemOpen
  \bibfield  {author} {\bibinfo {author} {\bibfnamefont {N.}~\bibnamefont
  {Brahms}}, \bibinfo {author} {\bibfnamefont {T.}~\bibnamefont {Botter}},
  \bibinfo {author} {\bibfnamefont {S.}~\bibnamefont {Schreppler}}, \bibinfo
  {author} {\bibfnamefont {D.~W.~C.}\ \bibnamefont {Brooks}}, \ and\ \bibinfo
  {author} {\bibfnamefont {D.~M.}\ \bibnamefont {Stamper-Kurn}},\ }\href
  {\doibase 10.1103/PhysRevLett.108.133601} {\bibfield  {journal} {\bibinfo
  {journal} {Phys. Rev. Lett.}\ }\textbf {\bibinfo {volume} {108}},\ \bibinfo
  {pages} {133601} (\bibinfo {year} {2012})}\BibitemShut {NoStop}%
\bibitem [{\citenamefont {Weinstein}\ \emph {et~al.}()\citenamefont
  {Weinstein}, \citenamefont {Lei}, \citenamefont {Wollman}, \citenamefont
  {Suh}, \citenamefont {Metelmann}, \citenamefont {Clerk},\ and\ \citenamefont
  {Schwab}}]{Weinstein14}%
  \BibitemOpen
  \bibfield  {author} {\bibinfo {author} {\bibfnamefont {A.~J.}\ \bibnamefont
  {Weinstein}}, \bibinfo {author} {\bibfnamefont {C.~U.}\ \bibnamefont {Lei}},
  \bibinfo {author} {\bibfnamefont {E.~E.}\ \bibnamefont {Wollman}}, \bibinfo
  {author} {\bibfnamefont {J.}~\bibnamefont {Suh}}, \bibinfo {author}
  {\bibfnamefont {A.}~\bibnamefont {Metelmann}}, \bibinfo {author}
  {\bibfnamefont {A.~A.}\ \bibnamefont {Clerk}}, \ and\ \bibinfo {author}
  {\bibfnamefont {K.~C.}\ \bibnamefont {Schwab}},\ }\href@noop {} {\ }\bibinfo
  {note} {ArXiv:1404.3242}\BibitemShut {NoStop}%
\bibitem [{\citenamefont {Marquardt}\ \emph {et~al.}(2007)\citenamefont
  {Marquardt}, \citenamefont {Chen}, \citenamefont {Clerk},\ and\ \citenamefont
  {Girvin}}]{Marquardt07}%
  \BibitemOpen
  \bibfield  {author} {\bibinfo {author} {\bibfnamefont {F.}~\bibnamefont
  {Marquardt}}, \bibinfo {author} {\bibfnamefont {J.~P.}\ \bibnamefont {Chen}},
  \bibinfo {author} {\bibfnamefont {A.~A.}\ \bibnamefont {Clerk}}, \ and\
  \bibinfo {author} {\bibfnamefont {S.~M.}\ \bibnamefont {Girvin}},\ }\href
  {\doibase 10.1103/PhysRevLett.99.093902} {\bibfield  {journal} {\bibinfo
  {journal} {Phys. Rev. Lett.}\ }\textbf {\bibinfo {volume} {99}},\ \bibinfo
  {pages} {093902} (\bibinfo {year} {2007})}\BibitemShut {NoStop}%
\bibitem [{\citenamefont {Wilson-Rae}\ \emph {et~al.}(2007)\citenamefont
  {Wilson-Rae}, \citenamefont {Nooshi}, \citenamefont {Zwerger},\ and\
  \citenamefont {Kippenberg}}]{WilsonRae07}%
  \BibitemOpen
  \bibfield  {author} {\bibinfo {author} {\bibfnamefont {I.}~\bibnamefont
  {Wilson-Rae}}, \bibinfo {author} {\bibfnamefont {N.}~\bibnamefont {Nooshi}},
  \bibinfo {author} {\bibfnamefont {W.}~\bibnamefont {Zwerger}}, \ and\
  \bibinfo {author} {\bibfnamefont {T.~J.}\ \bibnamefont {Kippenberg}},\ }\href
  {\doibase 10.1103/PhysRevLett.99.093901} {\bibfield  {journal} {\bibinfo
  {journal} {Phys. Rev. Lett.}\ }\textbf {\bibinfo {volume} {99}},\ \bibinfo
  {pages} {093901} (\bibinfo {year} {2007})}\BibitemShut {NoStop}%
\bibitem [{\citenamefont {Purdy}\ \emph {et~al.}(2013)\citenamefont {Purdy},
  \citenamefont {Peterson},\ and\ \citenamefont {Regal}}]{Purdy13}%
  \BibitemOpen
  \bibfield  {author} {\bibinfo {author} {\bibfnamefont {T.~P.}\ \bibnamefont
  {Purdy}}, \bibinfo {author} {\bibfnamefont {R.~W.}\ \bibnamefont {Peterson}},
  \ and\ \bibinfo {author} {\bibfnamefont {C.~A.}\ \bibnamefont {Regal}},\
  }\href {\doibase 10.1126/science.1231282} {\bibfield  {journal} {\bibinfo
  {journal} {Science}\ }\textbf {\bibinfo {volume} {339}},\ \bibinfo {pages}
  {801} (\bibinfo {year} {2013})}\BibitemShut {NoStop}%
\bibitem [{\citenamefont {Thompson}\ \emph {et~al.}(2008)\citenamefont
  {Thompson}, \citenamefont {Zwickl}, \citenamefont {Jayich}, \citenamefont
  {Marquardt}, \citenamefont {Girvin},\ and\ \citenamefont
  {Harris}}]{Thompson08}%
  \BibitemOpen
  \bibfield  {author} {\bibinfo {author} {\bibfnamefont {J.~D.}\ \bibnamefont
  {Thompson}}, \bibinfo {author} {\bibfnamefont {B.~M.}\ \bibnamefont
  {Zwickl}}, \bibinfo {author} {\bibfnamefont {A.~M.}\ \bibnamefont {Jayich}},
  \bibinfo {author} {\bibfnamefont {F.}~\bibnamefont {Marquardt}}, \bibinfo
  {author} {\bibfnamefont {S.~M.}\ \bibnamefont {Girvin}}, \ and\ \bibinfo
  {author} {\bibfnamefont {J.~G.~E.}\ \bibnamefont {Harris}},\ }\href {\doibase
  10.1038/nature06715} {\bibfield  {journal} {\bibinfo  {journal} {Nature}\
  }\textbf {\bibinfo {volume} {452}},\ \bibinfo {pages} {72} (\bibinfo {year}
  {2008})}\BibitemShut {NoStop}%
\bibitem [{\citenamefont {Purdy}\ \emph {et~al.}(2012)\citenamefont {Purdy},
  \citenamefont {Peterson}, \citenamefont {Yu},\ and\ \citenamefont
  {Regal}}]{Purdy12}%
  \BibitemOpen
  \bibfield  {author} {\bibinfo {author} {\bibfnamefont {T.~P.}\ \bibnamefont
  {Purdy}}, \bibinfo {author} {\bibfnamefont {R.~W.}\ \bibnamefont {Peterson}},
  \bibinfo {author} {\bibfnamefont {P.-L.}\ \bibnamefont {Yu}}, \ and\ \bibinfo
  {author} {\bibfnamefont {C.~A.}\ \bibnamefont {Regal}},\ }\href
  {http://stacks.iop.org/1367-2630/14/i=11/a=115021} {\bibfield  {journal}
  {\bibinfo  {journal} {New J. Phys.}\ }\textbf {\bibinfo {volume} {14}},\
  \bibinfo {pages} {115021} (\bibinfo {year} {2012})}\BibitemShut {NoStop}%
\bibitem [{\citenamefont {Yu}\ \emph {et~al.}(2014)\citenamefont {Yu},
  \citenamefont {Cicak}, \citenamefont {Kampel}, \citenamefont {Tsaturyan},
  \citenamefont {Purdy}, \citenamefont {Simmonds},\ and\ \citenamefont
  {Regal}}]{Yu14}%
  \BibitemOpen
  \bibfield  {author} {\bibinfo {author} {\bibfnamefont {P.-L.}\ \bibnamefont
  {Yu}}, \bibinfo {author} {\bibfnamefont {K.}~\bibnamefont {Cicak}}, \bibinfo
  {author} {\bibfnamefont {N.~S.}\ \bibnamefont {Kampel}}, \bibinfo {author}
  {\bibfnamefont {Y.}~\bibnamefont {Tsaturyan}}, \bibinfo {author}
  {\bibfnamefont {T.~P.}\ \bibnamefont {Purdy}}, \bibinfo {author}
  {\bibfnamefont {R.~W.}\ \bibnamefont {Simmonds}}, \ and\ \bibinfo {author}
  {\bibfnamefont {C.~A.}\ \bibnamefont {Regal}},\ }\href {\doibase
  http://dx.doi.org/10.1063/1.4862031} {\bibfield  {journal} {\bibinfo
  {journal} {Applied Physics Letters}\ }\textbf {\bibinfo {volume} {104}},\
  \bibinfo {pages} {023510} (\bibinfo {year} {2014})}\BibitemShut {NoStop}%
\bibitem [{\citenamefont {Tsaturyan}\ \emph {et~al.}(2014)\citenamefont
  {Tsaturyan}, \citenamefont {Barg}, \citenamefont {Simonsen}, \citenamefont
  {Villanueva}, \citenamefont {Schmid}, \citenamefont {Schliesser},\ and\
  \citenamefont {Polzik}}]{Tsaturyan14}%
  \BibitemOpen
  \bibfield  {author} {\bibinfo {author} {\bibfnamefont {Y.}~\bibnamefont
  {Tsaturyan}}, \bibinfo {author} {\bibfnamefont {A.}~\bibnamefont {Barg}},
  \bibinfo {author} {\bibfnamefont {A.}~\bibnamefont {Simonsen}}, \bibinfo
  {author} {\bibfnamefont {L.~G.}\ \bibnamefont {Villanueva}}, \bibinfo
  {author} {\bibfnamefont {S.}~\bibnamefont {Schmid}}, \bibinfo {author}
  {\bibfnamefont {A.}~\bibnamefont {Schliesser}}, \ and\ \bibinfo {author}
  {\bibfnamefont {E.~S.}\ \bibnamefont {Polzik}},\ }\href {\doibase
  10.1364/OE.22.006810} {\bibfield  {journal} {\bibinfo  {journal} {Opt.
  Express}\ }\textbf {\bibinfo {volume} {22}},\ \bibinfo {pages} {6810}
  (\bibinfo {year} {2014})}\BibitemShut {NoStop}%
\bibitem [{\citenamefont {Wilson}\ \emph {et~al.}(2009)\citenamefont {Wilson},
  \citenamefont {Regal}, \citenamefont {Papp},\ and\ \citenamefont
  {Kimble}}]{Wilson09}%
  \BibitemOpen
  \bibfield  {author} {\bibinfo {author} {\bibfnamefont {D.~J.}\ \bibnamefont
  {Wilson}}, \bibinfo {author} {\bibfnamefont {C.~A.}\ \bibnamefont {Regal}},
  \bibinfo {author} {\bibfnamefont {S.~B.}\ \bibnamefont {Papp}}, \ and\
  \bibinfo {author} {\bibfnamefont {H.~J.}\ \bibnamefont {Kimble}},\ }\href
  {\doibase 10.1103/PhysRevLett.103.207204} {\bibfield  {journal} {\bibinfo
  {journal} {Phys. Rev. Lett.}\ }\textbf {\bibinfo {volume} {103}},\ \bibinfo
  {pages} {207204} (\bibinfo {year} {2009})}\BibitemShut {NoStop}%
\bibitem [{\citenamefont {Weis}\ \emph {et~al.}(2010)\citenamefont {Weis},
  \citenamefont {Riviere}, \citenamefont {Deleglise}, \citenamefont {Gavartin},
  \citenamefont {Arcizet}, \citenamefont {Schliesser},\ and\ \citenamefont
  {Kippenberg}}]{Weis10}%
  \BibitemOpen
  \bibfield  {author} {\bibinfo {author} {\bibfnamefont {S.}~\bibnamefont
  {Weis}}, \bibinfo {author} {\bibfnamefont {R.}~\bibnamefont {Riviere}},
  \bibinfo {author} {\bibfnamefont {S.}~\bibnamefont {Deleglise}}, \bibinfo
  {author} {\bibfnamefont {E.}~\bibnamefont {Gavartin}}, \bibinfo {author}
  {\bibfnamefont {O.}~\bibnamefont {Arcizet}}, \bibinfo {author} {\bibfnamefont
  {A.}~\bibnamefont {Schliesser}}, \ and\ \bibinfo {author} {\bibfnamefont
  {T.~J.}\ \bibnamefont {Kippenberg}},\ }\href {\doibase
  10.1126/science.1195596} {\bibfield  {journal} {\bibinfo  {journal}
  {Science}\ }\textbf {\bibinfo {volume} {330}},\ \bibinfo {pages} {1520}
  (\bibinfo {year} {2010})}\BibitemShut {NoStop}%
\bibitem [{\citenamefont {Jayich}\ \emph {et~al.}(2012)\citenamefont {Jayich},
  \citenamefont {Sankey}, \citenamefont {Borkje}, \citenamefont {Lee},
  \citenamefont {Yang}, \citenamefont {Underwood}, \citenamefont {Childress},
  \citenamefont {Petrenko}, \citenamefont {Girvin},\ and\ \citenamefont
  {Harris}}]{Jayich12}%
  \BibitemOpen
  \bibfield  {author} {\bibinfo {author} {\bibfnamefont {A.~M.}\ \bibnamefont
  {Jayich}}, \bibinfo {author} {\bibfnamefont {J.~C.}\ \bibnamefont {Sankey}},
  \bibinfo {author} {\bibfnamefont {K.}~\bibnamefont {Borkje}}, \bibinfo
  {author} {\bibfnamefont {D.}~\bibnamefont {Lee}}, \bibinfo {author}
  {\bibfnamefont {C.}~\bibnamefont {Yang}}, \bibinfo {author} {\bibfnamefont
  {M.}~\bibnamefont {Underwood}}, \bibinfo {author} {\bibfnamefont
  {L.}~\bibnamefont {Childress}}, \bibinfo {author} {\bibfnamefont
  {A.}~\bibnamefont {Petrenko}}, \bibinfo {author} {\bibfnamefont {S.~M.}\
  \bibnamefont {Girvin}}, \ and\ \bibinfo {author} {\bibfnamefont {J.~G.~E.}\
  \bibnamefont {Harris}},\ }\href
  {http://stacks.iop.org/1367-2630/14/i=11/a=115018} {\bibfield  {journal}
  {\bibinfo  {journal} {New Journal of Physics}\ }\textbf {\bibinfo {volume}
  {14}},\ \bibinfo {pages} {115018} (\bibinfo {year} {2012})}\BibitemShut
  {NoStop}%
\bibitem [{\citenamefont {Pontin}\ \emph {et~al.}(2014)\citenamefont {Pontin},
  \citenamefont {Biancofiore}, \citenamefont {Serra}, \citenamefont
  {Borrielli}, \citenamefont {Cataliotti}, \citenamefont {Marino},
  \citenamefont {Prodi}, \citenamefont {Bonaldi}, \citenamefont {Marin},\ and\
  \citenamefont {Vitali}}]{Pontin14}%
  \BibitemOpen
  \bibfield  {author} {\bibinfo {author} {\bibfnamefont {A.}~\bibnamefont
  {Pontin}}, \bibinfo {author} {\bibfnamefont {C.}~\bibnamefont {Biancofiore}},
  \bibinfo {author} {\bibfnamefont {E.}~\bibnamefont {Serra}}, \bibinfo
  {author} {\bibfnamefont {A.}~\bibnamefont {Borrielli}}, \bibinfo {author}
  {\bibfnamefont {F.~S.}\ \bibnamefont {Cataliotti}}, \bibinfo {author}
  {\bibfnamefont {F.}~\bibnamefont {Marino}}, \bibinfo {author} {\bibfnamefont
  {G.~A.}\ \bibnamefont {Prodi}}, \bibinfo {author} {\bibfnamefont
  {M.}~\bibnamefont {Bonaldi}}, \bibinfo {author} {\bibfnamefont
  {F.}~\bibnamefont {Marin}}, \ and\ \bibinfo {author} {\bibfnamefont
  {D.}~\bibnamefont {Vitali}},\ }\href {\doibase 10.1103/PhysRevA.89.033810}
  {\bibfield  {journal} {\bibinfo  {journal} {Phys. Rev. A}\ }\textbf {\bibinfo
  {volume} {89}},\ \bibinfo {pages} {033810} (\bibinfo {year}
  {2014})}\BibitemShut {NoStop}%
\bibitem [{\citenamefont {Shkarin}\ \emph {et~al.}(2014)\citenamefont
  {Shkarin}, \citenamefont {Flowers-Jacobs}, \citenamefont {Hoch},
  \citenamefont {Kashkanova}, \citenamefont {Deutsch}, \citenamefont
  {Reichel},\ and\ \citenamefont {Harris}}]{Shkarin14}%
  \BibitemOpen
  \bibfield  {author} {\bibinfo {author} {\bibfnamefont {A.~B.}\ \bibnamefont
  {Shkarin}}, \bibinfo {author} {\bibfnamefont {N.~E.}\ \bibnamefont
  {Flowers-Jacobs}}, \bibinfo {author} {\bibfnamefont {S.~W.}\ \bibnamefont
  {Hoch}}, \bibinfo {author} {\bibfnamefont {A.~D.}\ \bibnamefont
  {Kashkanova}}, \bibinfo {author} {\bibfnamefont {C.}~\bibnamefont {Deutsch}},
  \bibinfo {author} {\bibfnamefont {J.}~\bibnamefont {Reichel}}, \ and\
  \bibinfo {author} {\bibfnamefont {J.~G.~E.}\ \bibnamefont {Harris}},\ }\href
  {\doibase 10.1103/PhysRevLett.112.013602} {\bibfield  {journal} {\bibinfo
  {journal} {Phys. Rev. Lett.}\ }\textbf {\bibinfo {volume} {112}},\ \bibinfo
  {pages} {013602} (\bibinfo {year} {2014})}\BibitemShut {NoStop}%
\bibitem [{\citenamefont {Khalili}\ \emph {et~al.}(2012)\citenamefont
  {Khalili}, \citenamefont {Miao}, \citenamefont {Yang}, \citenamefont
  {Safavi-Naeini}, \citenamefont {Painter},\ and\ \citenamefont
  {Chen}}]{Khalili12}%
  \BibitemOpen
  \bibfield  {author} {\bibinfo {author} {\bibfnamefont {F.~Y.}\ \bibnamefont
  {Khalili}}, \bibinfo {author} {\bibfnamefont {H.}~\bibnamefont {Miao}},
  \bibinfo {author} {\bibfnamefont {H.}~\bibnamefont {Yang}}, \bibinfo {author}
  {\bibfnamefont {A.~H.}\ \bibnamefont {Safavi-Naeini}}, \bibinfo {author}
  {\bibfnamefont {O.}~\bibnamefont {Painter}}, \ and\ \bibinfo {author}
  {\bibfnamefont {Y.}~\bibnamefont {Chen}},\ }\href {\doibase
  10.1103/PhysRevA.86.033840} {\bibfield  {journal} {\bibinfo  {journal} {Phys.
  Rev. A}\ }\textbf {\bibinfo {volume} {86}},\ \bibinfo {pages} {033840}
  (\bibinfo {year} {2012})}\BibitemShut {NoStop}%
\bibitem [{\citenamefont {Lee}\ \emph {et~al.}()\citenamefont {Lee},
  \citenamefont {Underwood}, \citenamefont {Mason}, \citenamefont {Shkarin},
  \citenamefont {Borkje}, \citenamefont {Girvin},\ and\ \citenamefont
  {Harris}}]{Lee14}%
  \BibitemOpen
  \bibfield  {author} {\bibinfo {author} {\bibfnamefont {D.}~\bibnamefont
  {Lee}}, \bibinfo {author} {\bibfnamefont {M.}~\bibnamefont {Underwood}},
  \bibinfo {author} {\bibfnamefont {D.}~\bibnamefont {Mason}}, \bibinfo
  {author} {\bibfnamefont {A.~B.}\ \bibnamefont {Shkarin}}, \bibinfo {author}
  {\bibfnamefont {K.}~\bibnamefont {Borkje}}, \bibinfo {author} {\bibfnamefont
  {S.~M.}\ \bibnamefont {Girvin}}, \ and\ \bibinfo {author} {\bibfnamefont
  {J.~G.~E.}\ \bibnamefont {Harris}},\ }\href@noop {} {\ }\bibinfo {note}
  {Private communication}\BibitemShut {NoStop}%
\end{thebibliography}
%

\end{document}